\documentclass[reprint,aps,prl]{revtex4-2}

\usepackage{amsmath}
\usepackage{amssymb}
\usepackage{graphicx}
\usepackage{dcolumn}
\usepackage{bm}
\usepackage{xcolor}
\usepackage{soul}
\usepackage{multirow}
\usepackage{booktabs}

\usepackage{notes2bib}

\begin{document}

\title{Hamiltonian Model for Energy Condensation in Classical Systems: Relevance to Proteins}

\author{Jordane Preto}
\affiliation{Computational Biomedicine (INM-9), Institute for Neuroscience and Medicine, Forschungszentrum Jülich, 52428 Jülich, Germany}
\affiliation{Aix-Marseille University, Université de Toulon, CNRS, Centre de Physique Théorique UMR 7332, 13288 Marseille Cedex 09, France}

\author{Vania Calandrini}
\affiliation{Computational Biomedicine (INM-9), Institute for Neuroscience and Medicine, Forschungszentrum Jülich, 52428 Jülich, Germany}

\author{Elena Floriani}
\affiliation{Aix-Marseille University, Université de Toulon, CNRS, Centre de Physique Théorique UMR 7332, 13288 Marseille Cedex 09, France}

\author{Gergely Katona}
\affiliation{Department of Chemistry and Molecular Biology, University of Gothenburg, 405 30 Gothenburg, Sweden}

\author{Marco Pettini}
\affiliation{Aix-Marseille University, Université de Toulon, CNRS, Centre de Physique Théorique UMR 7332, 13288 Marseille Cedex 09, France}
\affiliation{Quantum Biology Lab, Howard University, 2400 6th St NW, 16 17 Washington, DC 20059, USA}

\date{\today}

\begin{abstract}
Recent experimental evidence for collective protein vibrations in the terahertz (THz) domain indicates that energy in biomolecular systems can self-organize in an orderly manner, as anticipated by Fröhlich’s theory of condensates within a quantum framework. As a first step to bridge  THz experiments with theory, we study the Hamiltonian dynamics of a classical network of coupled normal modes representing Fröhlich-type systems. Our results demonstrate that biologically relevant condensates can emerge at room temperature under appropriate nonlinear coupling schemes. The condensation mechanism remains robust also when the original Fröhlich resonance conditions are relaxed.
\end{abstract}

\maketitle

\textit{Introduction}---Biological systems have always been known for their ability to self-regulate and adapt to external conditions. It is now widely accepted that such properties are the result of a long evolutionary process governed
by natural selection that promotes the proper functioning and longevity of organisms \cite{carroll2013dna}. As a result, biological structures often display energetic properties that appear extraordinary when compared to inert matter. 

Recent experiments have shown that spectra of proteins like bovine serum albumin (BSA) or R-phycoerythrin (R-PE) are not always thermalized but exhibit sharp sub-THz peaks upon laser light excitation \cite{nardecchia2018out,lechelon2022experimental}. In the case of R-PE, such low-frequency modes could be triggered not only upon (local) electronic excitation of protein fluorochromes, but also upon a temperature increase \cite{perez2025}. This proves that a disordered and delocalized form of energy like heat, which should simply distribute among all the normal modes of the protein, is instead driven into a well-defined collective oscillation. 

These results seem consistent with an old theoretical model proposed by H. Fröhlich \cite{frohlich1968long}, suggesting that, due to nonlinear processes, energy supplied to a set of THz modes may be specifically channeled into the lowest frequency mode, the so-called \textit{energy condensation}. This effect could have profound implications for energy storage \cite{frohlich1968long}, long-range selective intermolecular forces \cite{perez2025,lechelon2022experimental,preto2015possible}, and cognition \cite{stuart1998quantum}.

Starting from the numerical integration of Fröhlich's rate equations (FREs) in the semiclassical and classical limits, Reimers et al. \cite{reimers2009weak} identified regions of the parameter space leading to "weak" and "strong" condensates, which were discussed with respect to their possible biological relevance. The authors have also  explored the quantum coherence of condensates by mapping the rate equations into a quantum  Hamiltonian framework.

Experimental results mentioned above have also motivated force-field-based simulations, but the latter neither reproduced the observed spectral excitations nor confirmed or refuted Fröhlich's theory \cite{azizi2023examining,tenenbaum2024energy}. Possible reasons for these discrepancies include the modeling of the energy input as well as the significant difference between simulation ($\leq 1 \ \mu s$) and experimental ($10^{0}-10^{2} \,\mathrm{s}$) timescales.

In this work, we introduce a generic model showing that energy condensation of Fröhlich type can occur from simple classical Hamiltonian dynamics. Using a system of coupled oscillators representing relevant protein modes, the nearby environment (bath) and an energy source, we demonstrate that condensation in the lowest-frequency mode of the protein emerges for specific types of nonlinear (order-three) coupling with the environment. Energy injection and dissipation is achieved by coupling one thermostat to the bath and another to the source at a higher temperature placing the protein in out-of-equilibrium conditions.

With this setup, we found that energy condensation persisted when coupling constants and/or friction coefficients were rescaled over two orders of magnitude, highlighting the robustness of the phenomenon. While the model does not aim at explaining the experimental results of specific proteins, it establishes the foundations to explore Fröhlich condensation in quasi-classical open systems, as biomolecules are generally modeled.  Furthermore, projection of force field-based trajectories in the normal mode space is routinely performed to dissect the atomistic contributions to the essential functional dynamics of proteins (mainly driven by low-frequency modes) \cite{kolossvary2024}. Using this approach, out-of-equilibrium force field-based trajectories of proteins subjected to an instantaneous external perturbation have been well described in terms of coupled normal modes-models \cite{moritsugu2000}, which have highlighted the relevant role of order-three coupling in the energy redistribution among protein modes.

In this paper, we first formulate the problem using Fröhlich’s rate equations in the classical limit. We then introduce the classical Hamiltonian model and numerically examine the emergence of condensates across different parameter sets and nonlinear coupling conditions.

\begin{figure}[ht]
\includegraphics[scale=0.2]{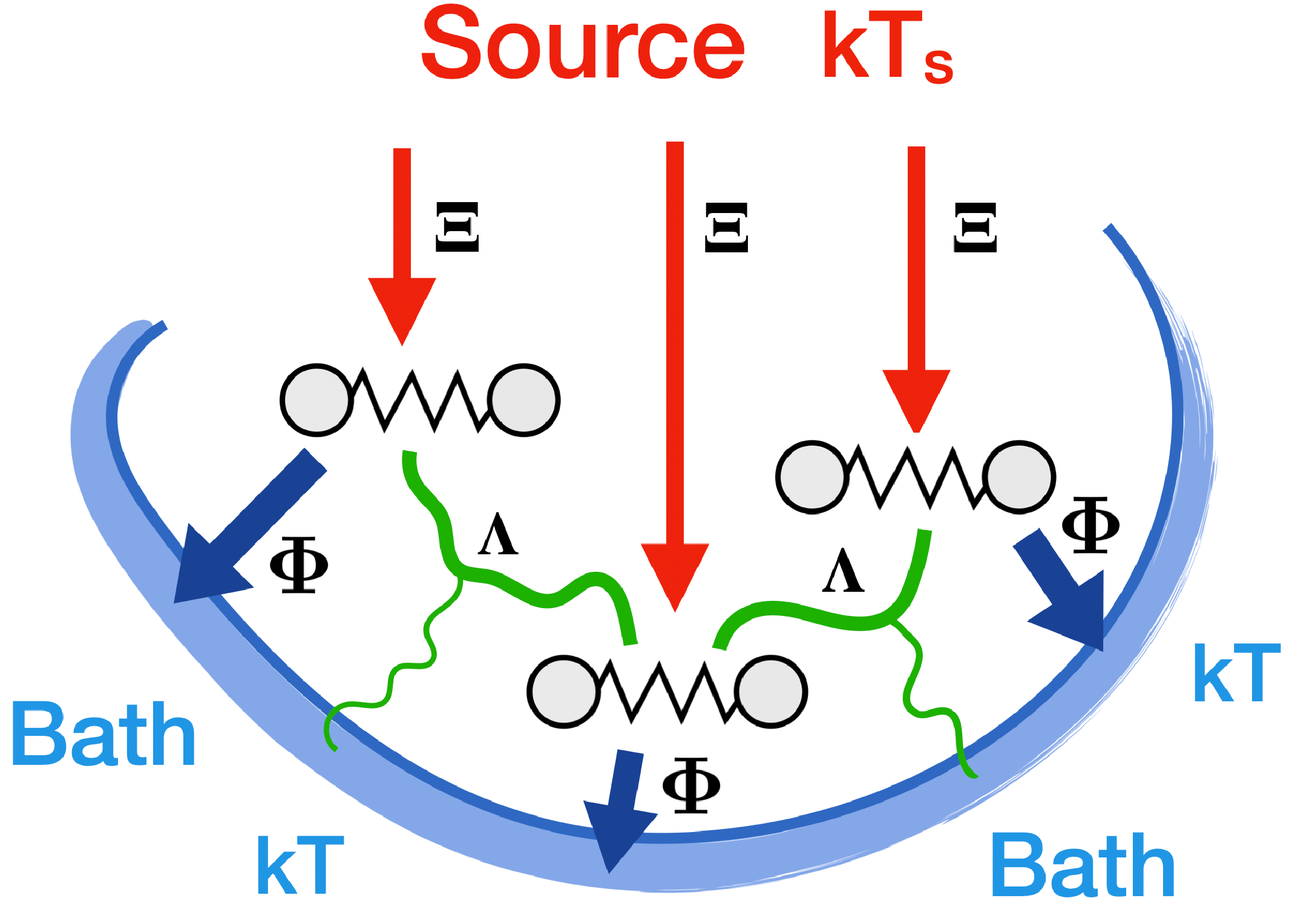}
\caption{\label{fig:epsart} Illustration of a Fröhlich system consisting of a set of oscillators, or modes, in contact with a heat bath at temperature $T$ and a source at temperature $T_S$. 
More details are given in the main text. Figure was inspired by Fig. 1 in \cite{reimers2009weak}.}
\end{figure}

\textit{Fröhlich systems}---A typical Fröhlich system is shown in Fig. \ref{fig:epsart} and consists of 3 components: (1) an ensemble of oscillators, or modes,
representing the protein of interest, (2) a heat bath representing the water/cell environment, and (3) an energy source (\textit{e.g.}, endogenous energy, light...). The protein modes interact
linearly with the heat bath and the source at rates $\Phi$ and $\Xi$, respectively, but they are also able to interact with each other. These nonlinear interactions are given by the $\Lambda$ rate and involve pairs of modes interacting via the thermal bath.

\begin{figure}[ht]
\includegraphics[scale=0.51]{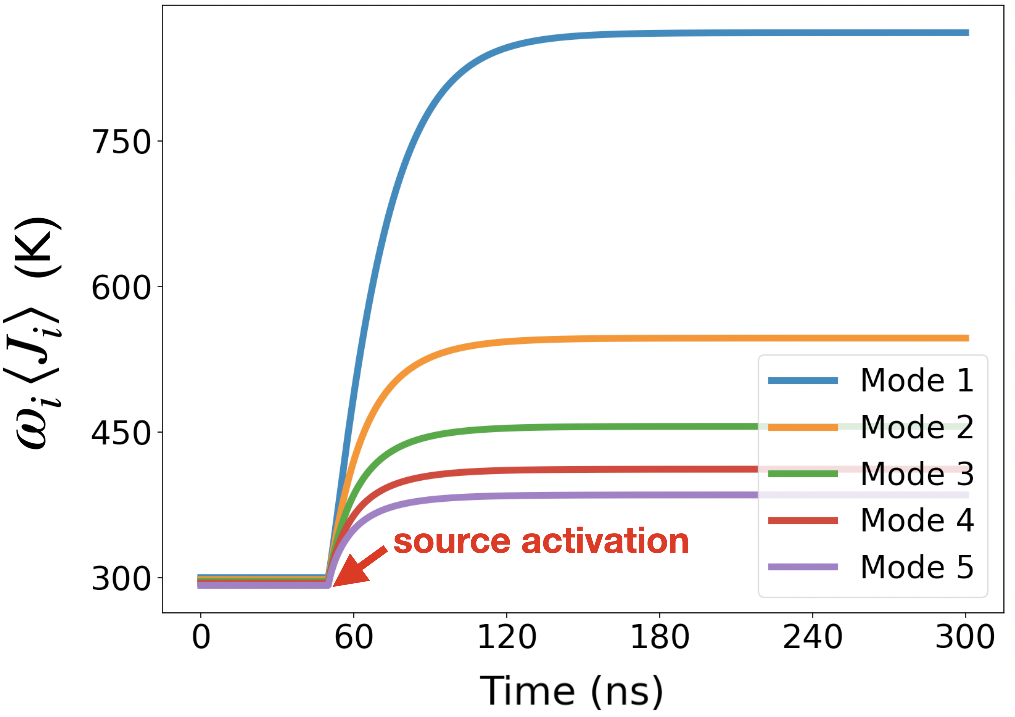}
\caption{\label{fig:epsart2} Time evolution of harmonic energies in a Fröhlich system made of $9$ protein modes; only the first 5 modes are shown.
Results were obtained by solving Eqs \eqref{eq:1} numerically with frequencies uniformly distributed ranging from $\omega_1 = 0.2\ \text {THz}$ to $\omega_9 = 1 \ \text {THz}$ with a $0.1$-THz increment
\cite{[While frequencies are typical of low-frequency modes in proteins (see for example ][){,} rate constant values were essentially chosen to show condensation within a time frame accessible to MD simulations.]go1983dynamics}.
Other parameters are $\Phi_i = 5\cdot10^{-5} \, \text{ps}^{-1}$, $\Lambda_{ij}= 5\cdot10^{-5} \, \text{ps}^{-1}$, $T=300 \ \text{K}$, $T_S=3000 \ \text{K}$.  $\Xi_i$ was set to $5\cdot10^{-6} \, \text{ps}^{-1}$ 
at $50$ ns ($\Xi_i = 0$ before $50$ ns).}
\vspace{-0.2cm}
\end{figure}

Based on the above assumptions, Fröhlich suggested a set of rate equations (FREs) to describe the energy of each protein mode. Calling $\langle J_i \rangle$ the average action of mode $i$ with 
frequency $\omega_i$, the FREs are given in the classical limit by \cite{lechelon2022experimental,preto2016classical}:

\begin{equation} \label{eq:1}
    \begin{split}
    \dot{\langle J_i \rangle}  & =  \Phi_i \left( \frac{kT}{\omega_i}  - \langle J_i \rangle \right) +  \Xi_i \left( \frac{kT_S}{ \omega_i}  - \langle J_i \rangle \right) + \\
 & \qquad \sum \limits_{j=1}^N \Lambda_{ij} \left (\langle J_j \rangle -   \langle J_i  \rangle   + \frac{ \omega_j - \omega_i}{kT}  \langle J_i \rangle  \langle J_j \rangle \right) \\
& \qquad \qquad  \qquad    \qquad  \mathrm{with} \ i = 1 \dots N,
   \end{split}
\end{equation}

where $N$ is the number of modes and the RHS includes the 3 coupling types introduced above: $\Phi$ and $\Xi$ (linear) and $\Lambda$ (nonlinear).
Remarkably, each term was originally postulated from the condition that energies $E_{i} = \omega_i \langle J_i \rangle$ are always equal to $kT$ or $kT_S$ 
in the stationary state, depending on whether the bath or the source is involved \cite{frohlich1968long}. For instance, if only the first term in the RHS is considered, $\dot{\langle J_i \rangle} = 0$ will give $\omega_i \langle J_i \rangle = kT$ for all $i$. The relationship between our rate constants and those of the original Fröhlich’s equations is given in the Supplemental Material \cite{suppmat}.  

\smallskip

Two important properties of Fröhlich systems can be deduced from the FREs. First, switching off nonlinear interactions, \textit{i.e.}, $\Lambda_{ij} = 0$, always leads to energy equipartition regardless of whether the source is active or not. This is 
illustrated in Fig. S1 in \cite{suppmat}, where the energy source is activated at $50 \ \text{ns}$. Secondly, if $\Lambda_{ij}$ is sufficiently large, energy will be channeled into the lowest frequency mode, provided that energy is supplied at a high enough rate $\Xi_i$. This phenomenon, known as Fröhlich condensation, is depicted in Fig. \ref{fig:epsart2} and was originally suggested to explain the emergence of specific low-frequency modes in biomolecular structures.

\textit{Hamiltonian dynamics}---A class of quantum Hamiltonians was proposed by Wu and Austin \cite{wu1977bose,wu1981frohlich} to describe the microscopic dynamics of Fröhlich systems. 
In this formalism, both the heat bath and the source are modelled as two additional sets of modes interacting with the protein. We focus here on the classical version of these Hamiltonians, given by $H =  H_0 + H_{int}$ where

\vspace{-0.4cm}
 
\begin{equation} \label{eq:2}
    \begin{split}
 H_0 =  \sum \limits_{i=1}^{N} & \frac{p_i^2}{2m_i} +  \frac{1}{2} m_i \omega_i^2 q_i^2  \: + \\        
  & \quad   \sum \limits_{k=1}^{N_B} \frac{p_k^{{\scriptscriptstyle (B)} 2}}{2m_k^{\scriptscriptstyle{(B)}}} + \frac{1}{2} m_k^{\scriptscriptstyle{(B)}} {\omega_k^{{\scriptscriptstyle (B)}  2}} q_k^{{\scriptscriptstyle (B)} 2}  \: + \\
   & \qquad   \qquad  \sum \limits_{l=1}^{N_S} \frac{p_l^{{\scriptscriptstyle (S)} 2}}{2m_l^{\scriptscriptstyle{(S)}}} + \frac{1}{2}  m_l^{\scriptscriptstyle (S)} {\omega_l^{{\scriptscriptstyle (S)}  2}} q_l^{{\scriptscriptstyle (S)} 2},
    \end{split}
\end{equation}

and $H_{int}$ is the interaction Hamiltonian such that

\vspace{-0.1cm}

\begin{equation}\label{eq:3}
H_{int}   =   \sum \limits_{ik} \phi_{ik} q_i q_k^{{\scriptscriptstyle (B)}}  + \sum \limits_{ik} \xi_{il} q_i q_l^{{\scriptscriptstyle (S)}}  + \sum \limits_{ijk} \lambda_{ijk} q_i q_j q_k^{{\scriptscriptstyle (B)}}.
\end{equation}

Here $p_i$, $p_k^{\scriptscriptstyle (B)}$,  $p_l^{\scriptscriptstyle (S)}$ and $q_i$, $q_k^{\scriptscriptstyle (B)}$,  $q_l^{\scriptscriptstyle (S)}$ are the momenta and the positions of the protein modes,
the bath and the source, respectively, while $m_i$,  $m_k^{\scriptscriptstyle{(B)}}$, $m_l^{\scriptscriptstyle{(S)}}$ and  $\omega_i$,  $\omega_k^{\scriptscriptstyle{(B)}}$, $\omega_l^{\scriptscriptstyle{(S)}}$ are their associated masses and frequencies. 
$N$, $N_B$ and $N_S$ are the numbers of modes in each set. Finally, $\phi_{ik}$, $\xi_{il}$ and $\lambda_{ijk}$ are the coupling coefficients related to the 3 types of interaction originally introduced by Fröhlich.

\smallskip

Previously, we have shown how the FREs can be recovered from a Hamiltonian similar to the one above \cite{preto2016classical}. This derivation supposes that the heat bath and the source are kept at temperature $T$ and $T_S$, respectively, and that 
nonlinear interactions are primarily driven by low-frequency modes of the bath. The latter assumption implies that resonances of the type $\omega_i + \omega_j - \omega_k^{\scriptscriptstyle{(B)}} = 0$ are negligible
over resonances of the type $\omega_i - \omega_j \pm \omega_k^{\scriptscriptstyle{(B)}} = 0$. 
Moreover, most derivations of the FREs are based on perturbation theory assuming high-order perturbative terms can either be included in the rate constants \cite{wu1981frohlich}, or are negligible \cite{mills1983frohlich}. 
This contrasts with many studies on rate equations showing both quantitative and qualitative changes when higher-order terms are considered \cite{zwanzig1972memory,nordholm1975systematic}. When using perturbation theory up to second order,
rate constants are related to $\phi_{ik}$, $\xi_{il}$ and $\lambda_{ijk}$ as follows \cite{preto2016classical}

\vspace{-0.2cm}

\begin{subequations} \label{eq:4}
 \begin{align}
& \Phi_i  =   \sum_{k} \frac{\alpha \, \phi_{ik}^2}{m_i m_k^{\scriptscriptstyle{(B)}}\omega_i^2} \, \delta (\omega_i - \omega_k^{\scriptscriptstyle (B)}) \\[10pt]
&\Xi_i   =    \sum_{l} \frac{\alpha \, \xi_{il}^2}{m_i m_l^{\scriptscriptstyle{(S)}}\omega_i^2} \, \delta (\omega_i - \omega_l^{\scriptscriptstyle (S)} ) \\[10pt]
& \Lambda_{ij}  =   \sum_{k}  \frac{2\alpha \, \lambda_{ijk}^2}{m_i m_j m_k^{\scriptscriptstyle{(B)}}}  \frac{kT \delta (\omega_i - \omega_j \pm \omega_k^{\scriptscriptstyle (B)} ) }{\omega_i \omega_j  (\omega_j - \omega_i)^2}
 \end{align}
\end{subequations}

where $\alpha$ is a time scaling factor to be determined. Note that $\Phi_i$, $\Xi_i$ and $\Lambda_{ij}$ depend on resonance conditions meaning that off-resonance interactions are supposed to have little impact on the dynamics.

\begin{figure}
\includegraphics[scale=0.35]{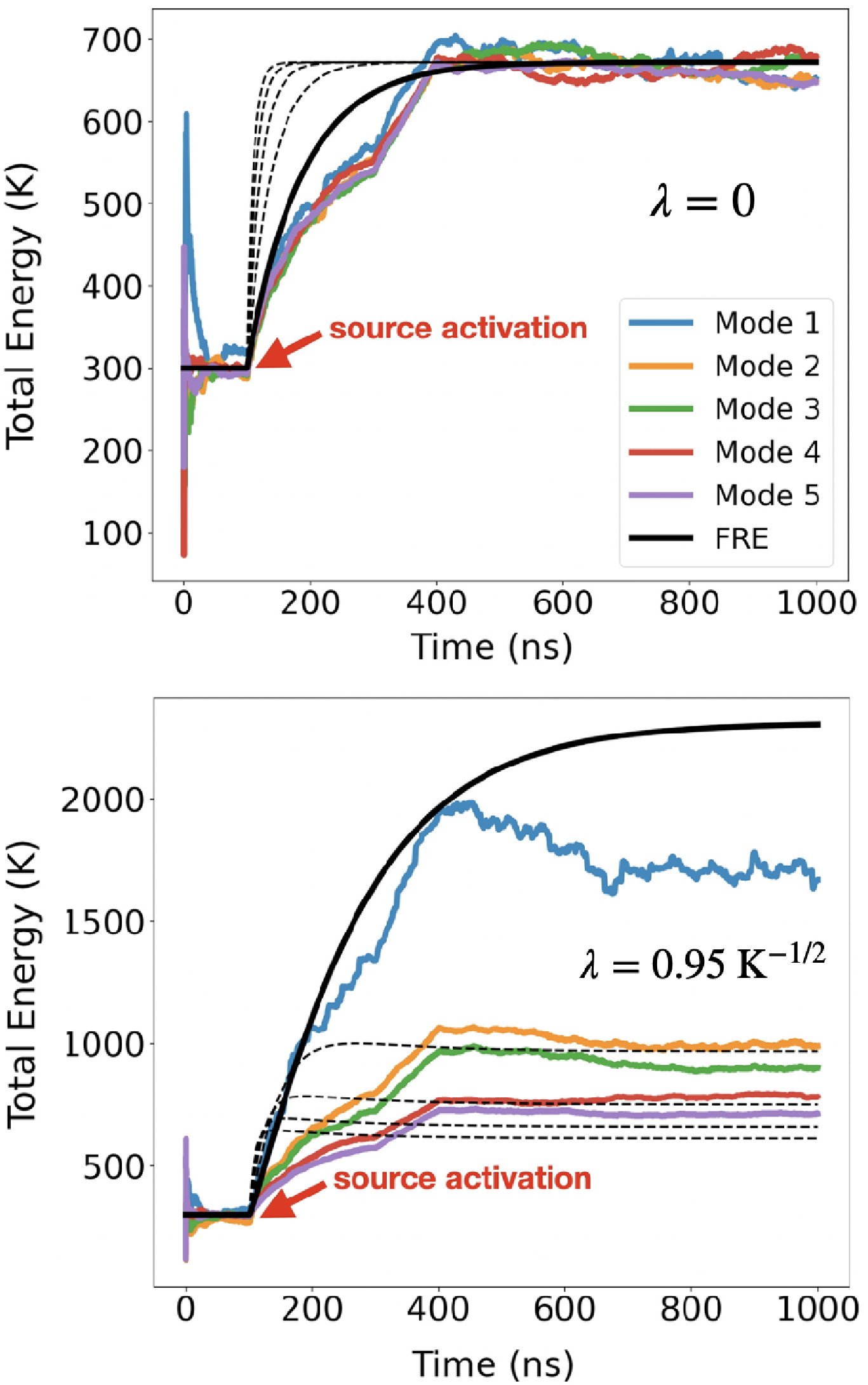}
\caption{\label{fig:epsart3} Time evolution of the total energies, \textit{i.e.}, the sum of kinetic and potential energies, in a Fröhlich system made of $9$ protein modes; only the first 5 modes are shown. 
Energies were computed as moving averages over $300$ ns. Results were obtained from Hamiltonian dynamics by keeping the bath at $T=300 \ \text{K}$ and the source at $T_S=3000 \ \text{K}$.
Protein frequencies were set uniformly from $\omega_1 = 0.2$ to $\omega_9 = 1$ THz with $0.1$-THz increment.  $\phi_{ik}$, $\xi_{il}$ and $\lambda_{ijk}$ were taken from Eqs. \eqref{eq:5}  with $\phi=1.0$ and 
$\xi=0.4$ (from $100$ ns). Top:  $\lambda=0.0$, bottom: $\lambda=0.95 \ \text{K}^{-1/2}$. Masses were all set to unity. Friction coefficients for the bath and source thermostats were both set to 0.1 $\mathrm{ps}^{-1}$. Curves in black correspond to the predictions of the FREs using Eqs. \eqref{eq:4} with $\alpha=0.02$ ps; 
the solid black line shows mode 1 while dashed curves correspond to secondary modes. Predictions of FREs are similar to those expected based on ref. \cite{reimers2009weak}}.
\vspace{-0.2cm}
\end{figure}

\smallskip

\textit{Numerical simulations}---To check the validity of the FREs and of the condensation effect, Hamilton's equations were integrated from $H_0+H_{int}$ given by  \eqref{eq:2} and \eqref{eq:3}. A velocity Verlet integration 
scheme was applied. The bath oscillators were kept at $T=300 \, \text{K}$ by coupling them to a Langevin thermostat while another Langevin thermostat was used to maintain the source oscillators at $T_S$, here treated as a free parameter. Bath and source thermostats were tested individually. This is shown in Fig. S2 in  \cite{suppmat}, where each set exhibits the right temperature as computed from ensemble and
time averages.

\smallskip

To get closer to Fröhlich's settings,
simulations were run by applying strict resonances only, \textit{i.e.}, coefficients $\phi_{ik}$, $\xi_{il}$ and $\lambda_{ijk}$ were all set to zero except at resonance conditions implied in Eqs. \eqref{eq:4}. Coupling coefficients were also supposed 
to be proportional to the square root of the masses and to the frequencies. For instance, in the case of $\phi$ coupling, the following coefficients were used when $\omega_i = \omega_k^{\scriptscriptstyle (B)}$:

\vspace{-0.2cm}

\begin{subequations} \label{eq:5}
\begin{align}
&\phi_{ik} = \phi \sqrt{m_i m_k^{\scriptscriptstyle{(B)}}} \omega_i \omega_k^{\scriptscriptstyle (B)} /N_\phi, \label{eq:5a} \\
\intertext{with $\phi_{ik}=0$ otherwise. Here $\phi$ is a unitless parameter and $N_\phi$ is the number of resonances related to $\phi$ interactions. A similar expression was used for $\xi_{il}$:}
&\indent\xi_{il} = \sqrt{m_i m_l^{\scriptscriptstyle{(S)}}} \omega_i \omega_l^{\scriptscriptstyle (S)}/N_\xi \label{eq:5b} \\
\intertext{when $\omega_i = \omega_l^{\scriptscriptstyle (S)}$ ($\xi_{il}=0$ otherwise). Finally, coefficients related to $\lambda$-coupling were set as}
&\lambda_{ijk} = \lambda \sqrt{m_i m_j m_k^{\scriptscriptstyle{(B)}}} \omega_i  \omega_j \omega_k^{\scriptscriptstyle (B)} /N_\lambda \label{eq:5c}
\end{align}
\end{subequations}

when $\omega_i-\omega_j \pm \omega_k^{\scriptscriptstyle (B)}=0$ ($\lambda_{ijk}=0$ otherwise) with $\lambda$ given in $\text{K}^{-1/2}$ units. In all our simulations, the bath was composed of $500$ oscillators with frequencies uniformly distributed from $0.002$ to $1.0$  THz, with increments of $0.002$ THz, while the source consisted of $100$ oscillators uniformly distributed from $0.01$ to $1.0$ THz, with increments of $0.01$ THz. Physical intuition for setting $\phi_{ik}$, $\xi_{il}$ and $\lambda_{ijk}$ from Eqs. \eqref{eq:5} is given in the next sections.

Results of our simulations are displayed in Fig. \ref{fig:epsart3}, where we reproduced the two main features of Fröhlich systems, that is, energy equipartition when nonlinear interactions are turned off ($\lambda = 0$), and condensation in the lowest frequency mode at high $\lambda$ value. Although the FREs show reasonable agreement with real dynamics, they also tend to miscalculate the energy
available in condensates, either by overestimating (Fig. \ref{fig:epsart3} bottom) or underestimating it (Fig. S3 in \cite{suppmat}), showing the limits of Eqs. \eqref{eq:1} combined with Eq. \eqref{eq:4}. 
Dynamics also revealed the formation of even more sizable condensates in other regions of the parameter space characterized by reasonable source and condensation energies (Fig. S3 in \cite{suppmat}). Energies in the range of $10^3 -10^4$ K equivalent temperature actually correspond to about 2-20 kcal/mol, which is the typical order of magnitude of the energy barrier associated with functional conformational rearrangements in proteins. 
Interestingly, in light-activated proteins 
energy conversion occurring within the protein upon light illumination leads to local and global structural rearrangements \cite{Burgie2014,Poddar2022}, which is compatible with the availability of a quite spatially delocalized and coherent form of energy. Phytochrome B photoreceptor exploits temperature to safely reverse from active to inactive conformational state \cite{Legris2016}, which proves how proteins can use a disordered and incoherent form of energy to accomplish a well-defined collective rearrangement.

We note here that using the index and threshold value introduced in \cite{reimers2009weak}, the observed condensates would be classified as "weak". 
Based on the energetic argument above, the threshold given in \cite{reimers2009weak} to discriminate biologically relevant/not relevant energies can be relaxed.

\begin{figure}[ht]
\includegraphics[scale=0.34]{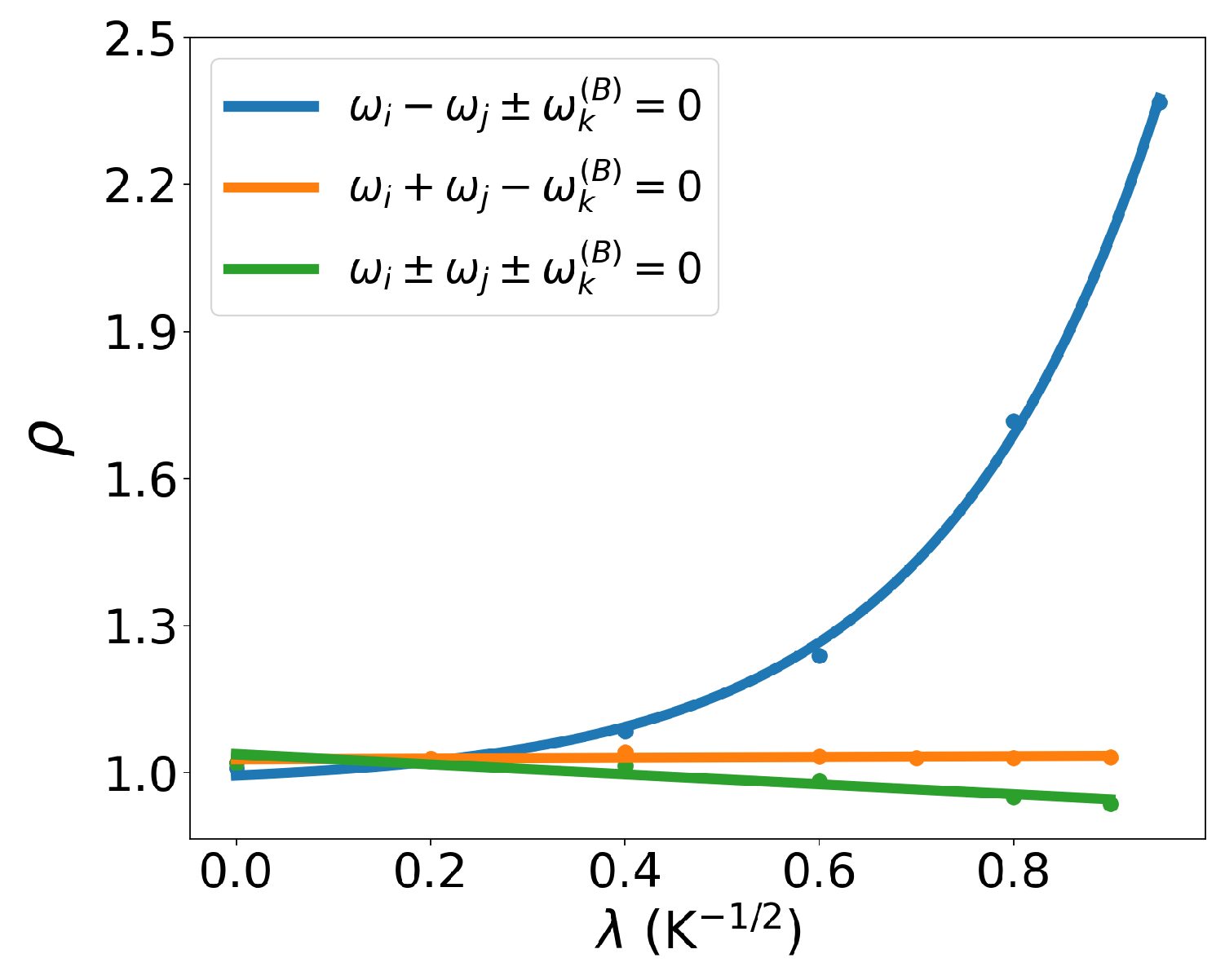}
\caption{\label{fig:epsart4} Condensation index $\rho$ of a Fröhlich system as deduced from Hamiltonian dynamics for different $\lambda$ values.
$\lambda_{ijk}$ coefficients were calculated from Eq. \eqref{eq:5c} under different types of resonance: $\omega_i - \omega_j  \pm \omega_k^{\scriptscriptstyle (B)} = 0$ (Fröhlich),  $\omega_i + \omega_j  - \omega_k^{\scriptscriptstyle (B)} = 0$ (Lifshits) 
and $\omega_i \pm \omega_j  \pm \omega_k^{\scriptscriptstyle (B)} = 0$ (combination of both). Other parameters remain the same as Fig. \ref{fig:epsart3}.}
\vspace{-0.4cm}
\end{figure}
\smallskip

\textit{Other coupling schemes}--- Our choice to apply Eq. \eqref{eq:5c} only when $\omega_i - \omega_j  \pm \omega_k^{\scriptscriptstyle (B)} = 0$ (Fröhlich resonances) was motivated by the fact that the FREs implicitly assume that such nonlinear resonances
are predominant over other types of resonance like $\omega_i + \omega_j  - \omega_k^{\scriptscriptstyle (B)} = 0$. Lifshits \cite{Lifshits1972problem} identified the latter type as a possible hindrance to condensation. To investigate this, another round of simulations was run by computing the $\lambda_{ijk}$ coefficients from \eqref{eq:5c} 
only when $\omega_i + \omega_j  - \omega_k^{\scriptscriptstyle (B)} = 0$ ($\lambda_{ijk}=0$ otherwise). Extra simulations were performed in the case $\omega_i \pm \omega_j  \pm \omega_k^{\scriptscriptstyle (B)} = 0$, \textit{i.e.}, when both Fröhlich and Lifshits resonances were involved. 

Results are depicted in Fig. \ref{fig:epsart4} highlighting the condensation ratio $\rho$ as a function of the $\lambda$ parameter. 
Here $\rho$ was defined as the energy of mode 1 in the stationary state divided by the average energy over the secondary modes, \textit{i.e.},  $\rho = E_1/\langle E_i \rangle_{i=2..N}$. Thus, $\rho=1$ 
indicates energy equipartition whereas $\rho>1$ implies condensation. From Fig. \ref{fig:epsart4}, we see that resonances involving high-frequency modes of the bath tend to destroy the condensation effect, regardless of whether Fröhlich resonances are included or not. Although it was suggested that the magnitude of Lifshits resonances is likely to be negligible in practice \cite{mills1983frohlich}, investigation of Fröhlich condensates in real structures would require careful evaluation of these contributions as deduced from all-atom force fields.

Moving further away from the ideal Fröhlich resonance conditions, we also investigated the regime in which the resonance constraints on both linear and nonlinear coupling terms are fully relaxed. In this scenario, $\phi_{ik}$ , $\xi_{il}$ and $\lambda_{ijk}$ were set from Eqs. \eqref{eq:5a}, \eqref{eq:5b} and \eqref{eq:5c}, respectively, for all $i,k$ and $l$. In this case, no condensate forms in the dynamics, possibly owing to Lifshits resonances that could interfere with the condensation mechanism.

Since Lifshits resonances typically involve higher bath frequencies than Fröhlich ones, nonlinear interactions with a low-pass bath were tested. Specifically, the $\lambda_{ijk}$'s were computed from \eqref{eq:5c} only when $\omega_k^{\scriptscriptstyle (B)} \leq\omega_0^{\scriptscriptstyle (B)}$ ($\lambda_{ijk}=0$ otherwise). Here $\omega_0^{\scriptscriptstyle (B)}$ refers to a typical bath frequency beyond which Lifshits resonances become dominant over Fröhlich ones (Fig. S4 in \cite{suppmat}). This type of nonlinear coupling was able to restore energy condensation (Fig. S5 in \cite{suppmat}), suggesting that low bath frequencies are crucial for redistributing energy into the lowest-frequency protein mode (see also Table \ref{tab:1} for a summary of conditions
leading to condensation).

\renewcommand{\arraystretch}{1.9}
\begin{table}[ht]
\setlength{\abovecaptionskip}{1pt}
\setlength{\tabcolsep}{7pt}
\caption{Resonance conditions leading to condensation; $\lambda_{ijk}$ is computed from Eq. \eqref{eq:5c} when the condition is satisfied and set to $0$ otherwise. In the first 3 cases, linear coefficients $\phi_{ik}$ and $\xi_{il}$ are computed from Eqs. \eqref{eq:5a} and \eqref{eq:5b} only when $\omega_i=\omega_k^{\scriptscriptstyle (B)}$ and $\omega_i=\omega_l^{\scriptscriptstyle (S)}$ and set to $0$ otherwise. For the last 2 cases, linear coefficients are computed from the same equations for all $i$, $k$, and $l$.}
\label{tab:1}
\vspace{1em}
\centering
\begin{tabular}{ccc}
\toprule
\textbf{Case} & \textbf{Condition on} $\lambda_{ijk}$ & \textbf{Condensation} \\
\midrule
Fröhlich  & $\omega_i - \omega_j \pm \omega_k^{(B)} = 0$ & {\large\textcolor{green}{$\checkmark$}} \\
Lifshits  & $\omega_i + \omega_j - \omega_k^{(B)} = 0$ & {\large \textcolor{red}{$\times$}} \\
All resonances & $\omega_i \pm \omega_j \pm \omega_k^{(B)} = 0$ & {\large \textcolor{red}{$\times$}} \\
Low-pass bath & $\omega_k^{\scriptscriptstyle (B)} \leq\omega_0^{\scriptscriptstyle (B)}$ & {\large\textcolor{green}{$\checkmark$}} \\
General & no restrictions & {\large \textcolor{red}{$\times$}} \\
\bottomrule
\end{tabular}
\end{table}

\smallskip

\textit{Protein-bath vs protein-protein coupling}---In addition to nonlinear resonances, we explored the importance of the heat bath in nonlinear coupling. To this purpose, 
we modified the cubic potential in Eq. \eqref{eq:3}, replacing the $\lambda_{ijk} q_i q_j q_k^{{\scriptscriptstyle (B)}}$ term with a
$\lambda_{ijk} q_i q_j q_k$ term involving only protein modes. This modification was sufficient to completely suppress Fr\"ohlich condensation
across all tested parameter regions where the phenomenon was originally observed. In all cases, energy equipartition was observed (not shown). 
While we cannot rule out the existence of Fr\"ohlich condensates in unexplored regions of the parameter space, our results strongly indicate that bath-mediated
coupling is essential for inducing condensation.

\smallskip

\textit{Constant coupling} ---Our simulations were run by setting $\phi_{ik}$, $\xi_{il}$ and $\lambda_{ijk}$ proportional to the square root of the masses and to the frequencies. Such a choice was motivated by the fact that, at thermal equilibrium, averaged potential energies $\frac{1}{2} m_i \omega_i^2 \langle q_i^2 \rangle$ should equal $kT/2$ or $kT_S/2$ for the bath or the energy source, respectively. Thus, 2 modes 
with the same mass but different frequencies, say $\omega_1< \omega_2$, should satisfy $\langle q_1^2 \rangle > \langle q_2^2 \rangle$ in order to preserve equilibrium condition, meaning that low-frequency modes will naturally have higher amplitude than high-frequency ones. In that case, an interaction potential with constant coefficients (\textit{e.g.}, $\phi_{ik} q_i q_k^{{\scriptscriptstyle (B)}}$ with $\phi_{ik}= \phi$) will generate stronger interactions between low-frequency modes. This observation can also be made for masses, \textit{i.e.}, modes with a large mass will have a stronger impact on the interaction energy. Setting coefficients as in Eqs  \eqref{eq:5} enables to circumvent this issue, ensuring that all the modes contribute equally to the potential.

Despite the above, the impact of a constant coupling on a Fr\"ohlich system was still investigated. 
In this scenario, our results showed that the interaction energy decreases exponentially as Fröhlich condensation takes place. 
More precisely, the interaction energy in the lowest frequency mode becomes of the same order of magnitude as the harmonic energy, then rapidly diverges into negative values (not shown).
This result is clearly not compatible with our Hamiltonian model whereby $H_{int}$ is always considered a small perturbation as compared to $H_0$. Again, using Eqs  \eqref{eq:5}
helped fix this issue by keeping the interaction energy reasonably low even when sizable condensation occurs (Fig. S6 in \cite{suppmat}). 

\smallskip

\textit{Thermostats}---Aside from nonlinear terms and the choice of coupling coefficients, thermostats used to maintain the bath and the energy source at the target temperature may 
also have a non-negligible impact on the emergence of condensation. 
The influence of the Langevin friction coefficient of the bath (\mbox{$\gamma$}) 
and the source (\mbox{$\gamma_S$}) was investigated by varying each individually (Table S1) or both simultaneously (Table S2) from 0.03 to 1 ps$^{-1}$, while keeping other parameters fixed. Results show that condensation is overall robust and its extent can be modulated by varying \mbox{$\gamma$} and/or \mbox{$\gamma_S$}, highlighting the importance of dissipation and injection rates on Fr\"ohlich effect.

\smallskip

\textit{Rescaling study}---To further assess the robustness of the condensation mechanism and the impact of the parameters on the timescale of the condensation,  
coupling coefficients were all rescaled by the same multiplicative factor $\kappa$ over two orders of magnitude. Parameters used to produce Fig.~\ref{fig:epsart3} were taken as a reference ($\kappa = 1.0$). The $\kappa$ factor was then varied from $0.01$ to $1$ and applied simultaneously to $\phi$, $\xi$, and $\lambda$, while the friction coefficients $\gamma$ and $\gamma_S$ are left unchanged. Fig. S7 in \cite{suppmat} shows the condensation ratio $\rho$ as a function of $\kappa$, along with the corresponding condensation time. 
The latter was obtained by fitting the average energy of the lowest-frequency mode as a function of time. Fitting was performed using the function $A-B\exp{(-t/\tau)}$ from the moment the source is activated ($\xi\neq0$) when the energy is at the equipartition value of 300~K. The condensation time was then defined as the fitted coefficient $\tau$ multiplied by $\ln(2)$ corresponding to the time to reach half of the plateau value. From our analysis, we see that $\rho$ strongly increases as $\kappa$ decreases, while the condensation time follows a similar trend but rises much more sharply at small $\kappa$ values. 

\begin{figure}[ht]
\centering
\includegraphics[scale=0.53]{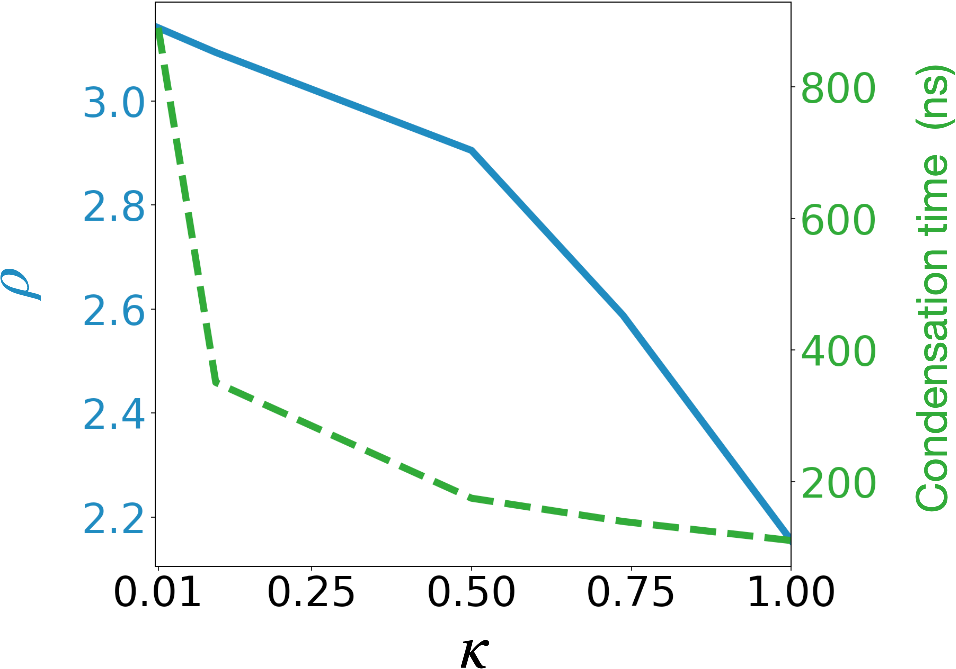}
\caption{Blue: condensation ratio $\rho$ as a function of the scaling factor $\kappa$ multiplying the coupling coefficients $\phi$, $\xi$, and $\lambda$ and the friction coefficients $\gamma$ and $\gamma_S$ simultaneously. $\kappa=1.0$ corresponds to the parameters used in Fig.~\ref{fig:epsart3}. Except for $\phi$, $\xi$, $\lambda$, $\gamma$, and $\gamma_S$, all other parameters are the same as in Fig.~\ref{fig:epsart3} for all $\kappa$ values. Green: condensation time (see main text for details).}
\label{fig:epsart5}
\vspace{-0.4cm}
\end{figure}

The same rescaling study was repeated by applying the $\kappa$ factor to both the friction coefficients $\gamma$ and $\gamma_S$, and to the coupling constants. Results are displayed in Fig.~\ref{fig:epsart5}. As in the previous case, a steep increase of the condensation time is observed as $\kappa$ becomes smaller. Although $\rho$ also increases, a clear concave dependence is found in this case, with $\rho$ stabilizing at very small $\kappa$. This highlights conditions in which the condensation time can grow to very large values while keeping the condensed energy relatively stable.

\smallskip

\textit{Discussion and perspectives}---The present study deals with the foundations of the condensation phenomenon proposed by Fr\"ohlich in the 1960s aimed at explaining coherent 
excitations in biomolecular systems \cite{frohlich1968long}. Motivated by recent experimental findings \cite{nardecchia2018out,lechelon2022experimental,perez2025}, we simulated the dynamics of a general classical 
Hamiltonian that incorporates all the key elements of Fr\"ohlich's models. Our results show that condensation, including strong condensates with reasonable energy values,
can emerge under specific types of nonlinear coupling (Fr\"ohlich resonances), supporting the physical plausibility of Fr\"ohlich's effect. 
Conversely, the addition of resonances 
involving high-frequency modes of the bath (Lifshits resonances) or the use of a nonlinear potential without bath mediation, result in suppressing the phenomenon. Interestingly, condensation still emerges without specific resonance conditions provided that nonlinear interactions are driven by low-frequency bath modes. This was implemented by introducing a low-pass filter on the bath frequencies, where the threshold is the frequency at which the number of verified Lifshits resonances exceeds that of Fröhlich's. These factors should be carefully examined when investigating the condensation effect in real biomolecular structures.

Given the large number of free parameters involved in the dynamics of Fr\"ohlich's systems, a full exploration of the
parameter space is beyond the scope of this letter. 
Although 
higher-order coupling mechanisms could bring more realistic characteristics to the model \cite{moritsugu2003temperature}, 
a more practical approach would be to tune the order-three coupling parameters so far considered directly from standard MD force fields targeting a specific model protein. Possible strategies include normal mode analysis (NMA) to generate the 
set of eigenvectors needed to transition from real space to the space of protein normal modes \cite{pandey2024tubulin,bauer2019normal}. In this case, not only normal frequencies of the protein and the heat bath could be deduced but eigenvectors may also be used to evaluate the force field function in the space of modes. This would allow coupling coefficients in Eq. \eqref{eq:3} to be deduced and used in our model to verify whether condensation can indeed arise. 
Furthermore, we note that energy condensation remains sizable for a quite large interval of source and bath friction coefficients, including typical values used in MD simulations. Finally, while our frequencies are representative of low-frequency modes in proteins, coupling constants values were essentially chosen to observe a well-formed condensate within a reasonable timescale, typically on the order of several hundreds of ns. Reducing the magnitude of the coupling constants in our Hamiltonian was found to delay the formation of condensates. Although at this stage one cannot draw a direct link between the timescales explored by a typical MD force field and those explored in our model, our findings suggest that the length of force field-based simulations could be a possible limiting factor in observing a clear signature of well-formed Fr\"ohlich's condensates.

\medskip

\textit{Acknowledgments}---This work and J. Preto’s postdoctoral fellowship were supported by the European Union’s Horizon Research and Innovation Programme under Grant Agreement No. 964203 (FET-Open LINkS project). 

\appendix

\setcounter{figure}{0}
\renewcommand{\figurename}{Fig.}
\renewcommand{\thefigure}{S\arabic{figure}}


\begin{thebibliography}{27}%
\makeatletter
\providecommand \@ifxundefined [1]{%
 \@ifx{#1\undefined}
}%
\providecommand \@ifnum [1]{%
 \ifnum #1\expandafter \@firstoftwo
 \else \expandafter \@secondoftwo
 \fi
}%
\providecommand \@ifx [1]{%
 \ifx #1\expandafter \@firstoftwo
 \else \expandafter \@secondoftwo
 \fi
}%
\providecommand \natexlab [1]{#1}%
\providecommand \enquote  [1]{``#1''}%
\providecommand \bibnamefont  [1]{#1}%
\providecommand \bibfnamefont [1]{#1}%
\providecommand \citenamefont [1]{#1}%
\providecommand \href@noop [0]{\@secondoftwo}%
\providecommand \href [0]{\begingroup \@sanitize@url \@href}%
\providecommand \@href[1]{\@@startlink{#1}\@@href}%
\providecommand \@@href[1]{\endgroup#1\@@endlink}%
\providecommand \@sanitize@url [0]{\catcode `\\12\catcode `\$12\catcode
  `\&12\catcode `\#12\catcode `\^12\catcode `\_12\catcode `\%12\relax}%
\providecommand \@@startlink[1]{}%
\providecommand \@@endlink[0]{}%
\providecommand \url  [0]{\begingroup\@sanitize@url \@url }%
\providecommand \@url [1]{\endgroup\@href {#1}{\urlprefix }}%
\providecommand \urlprefix  [0]{URL }%
\providecommand \Eprint [0]{\href }%
\providecommand \doibase [0]{https://doi.org/}%
\providecommand \selectlanguage [0]{\@gobble}%
\providecommand \bibinfo  [0]{\@secondoftwo}%
\providecommand \bibfield  [0]{\@secondoftwo}%
\providecommand \translation [1]{[#1]}%
\providecommand \BibitemOpen [0]{}%
\providecommand \bibitemStop [0]{}%
\providecommand \bibitemNoStop [0]{.\EOS\space}%
\providecommand \EOS [0]{\spacefactor3000\relax}%
\providecommand \BibitemShut  [1]{\csname bibitem#1\endcsname}%
\let\auto@bib@innerbib\@empty
\bibitem [{\citenamefont {Carroll}\ \emph {et~al.}(2013)\citenamefont
  {Carroll}, \citenamefont {Grenier},\ and\ \citenamefont
  {Weatherbee}}]{carroll2013dna}%
  \BibitemOpen
  \bibfield  {author} {\bibinfo {author} {\bibfnamefont {S.~B.}\ \bibnamefont
  {Carroll}}, \bibinfo {author} {\bibfnamefont {J.~K.}\ \bibnamefont
  {Grenier}},\ and\ \bibinfo {author} {\bibfnamefont {S.~D.}\ \bibnamefont
  {Weatherbee}},\ }\href@noop {} {\emph {\bibinfo {title} {From DNA to
  diversity: molecular genetics and the evolution of animal design}}}\
  (\bibinfo  {publisher} {John Wiley \& Sons},\ \bibinfo {year}
  {2013})\BibitemShut {NoStop}%
\bibitem [{\citenamefont {Nardecchia}\ \emph {et~al.}(2018)\citenamefont
  {Nardecchia}, \citenamefont {Torres}, \citenamefont {Lechelon}, \citenamefont
  {Giliberti}, \citenamefont {Ortolani}, \citenamefont {Nouvel}, \citenamefont
  {Gori}, \citenamefont {Meriguet}, \citenamefont {Donato}, \citenamefont
  {Preto}, \citenamefont {Varani}, \citenamefont {Sturgis},\ and\ \citenamefont
  {Pettini}}]{nardecchia2018out}%
  \BibitemOpen
  \bibfield  {author} {\bibinfo {author} {\bibfnamefont {I.}~\bibnamefont
  {Nardecchia}}, \bibinfo {author} {\bibfnamefont {J.}~\bibnamefont {Torres}},
  \bibinfo {author} {\bibfnamefont {M.}~\bibnamefont {Lechelon}}, \bibinfo
  {author} {\bibfnamefont {V.}~\bibnamefont {Giliberti}}, \bibinfo {author}
  {\bibfnamefont {M.}~\bibnamefont {Ortolani}}, \bibinfo {author}
  {\bibfnamefont {P.}~\bibnamefont {Nouvel}}, \bibinfo {author} {\bibfnamefont
  {M.}~\bibnamefont {Gori}}, \bibinfo {author} {\bibfnamefont {Y.}~\bibnamefont
  {Meriguet}}, \bibinfo {author} {\bibfnamefont {I.}~\bibnamefont {Donato}},
  \bibinfo {author} {\bibfnamefont {J.}~\bibnamefont {Preto}}, \bibinfo
  {author} {\bibfnamefont {L.}~\bibnamefont {Varani}}, \bibinfo {author}
  {\bibfnamefont {J.}~\bibnamefont {Sturgis}},\ and\ \bibinfo {author}
  {\bibfnamefont {M.}~\bibnamefont {Pettini}},\ }\bibfield  {title} {\bibinfo
  {title} {Out-of-equilibrium collective oscillation as phonon condensation in
  a model protein},\ }\href@noop {} {\bibfield  {journal} {\bibinfo  {journal}
  {Physical Review X}\ }\textbf {\bibinfo {volume} {8}},\ \bibinfo {pages}
  {031061} (\bibinfo {year} {2018})}\BibitemShut {NoStop}%
\bibitem [{\citenamefont {Lechelon}\ \emph {et~al.}(2022)\citenamefont
  {Lechelon}, \citenamefont {Meriguet}, \citenamefont {Gori}, \citenamefont
  {Ruffenach}, \citenamefont {Nardecchia}, \citenamefont {Floriani},
  \citenamefont {Coquillat}, \citenamefont {Teppe}, \citenamefont {Mailfert},
  \citenamefont {Marguet}, \citenamefont {Ferrier}, \citenamefont {Varani},
  \citenamefont {Sturgis}, \citenamefont {Torres},\ and\ \citenamefont
  {Pettini}}]{lechelon2022experimental}%
  \BibitemOpen
  \bibfield  {author} {\bibinfo {author} {\bibfnamefont {M.}~\bibnamefont
  {Lechelon}}, \bibinfo {author} {\bibfnamefont {Y.}~\bibnamefont {Meriguet}},
  \bibinfo {author} {\bibfnamefont {M.}~\bibnamefont {Gori}}, \bibinfo {author}
  {\bibfnamefont {S.}~\bibnamefont {Ruffenach}}, \bibinfo {author}
  {\bibfnamefont {I.}~\bibnamefont {Nardecchia}}, \bibinfo {author}
  {\bibfnamefont {E.}~\bibnamefont {Floriani}}, \bibinfo {author}
  {\bibfnamefont {D.}~\bibnamefont {Coquillat}}, \bibinfo {author}
  {\bibfnamefont {F.}~\bibnamefont {Teppe}}, \bibinfo {author} {\bibfnamefont
  {S.}~\bibnamefont {Mailfert}}, \bibinfo {author} {\bibfnamefont
  {D.}~\bibnamefont {Marguet}}, \bibinfo {author} {\bibfnamefont
  {P.}~\bibnamefont {Ferrier}}, \bibinfo {author} {\bibfnamefont
  {L.}~\bibnamefont {Varani}}, \bibinfo {author} {\bibfnamefont
  {J.}~\bibnamefont {Sturgis}}, \bibinfo {author} {\bibfnamefont
  {J.}~\bibnamefont {Torres}},\ and\ \bibinfo {author} {\bibfnamefont
  {M.}~\bibnamefont {Pettini}},\ }\bibfield  {title} {\bibinfo {title}
  {Experimental evidence for long-distance electrodynamic intermolecular
  forces},\ }\href@noop {} {\bibfield  {journal} {\bibinfo  {journal} {Science
  Advances}\ }\textbf {\bibinfo {volume} {8}},\ \bibinfo {pages} {eabl5855}
  (\bibinfo {year} {2022})}\BibitemShut {NoStop}%
\bibitem [{\citenamefont {Perez-Martin}\ \emph {et~al.}(2025)\citenamefont
  {Perez-Martin}, \citenamefont {Beranger}, \citenamefont {Bonnet},
  \citenamefont {Teppe}, \citenamefont {Lisauskas}, \citenamefont {Ikamas},
  \citenamefont {Vrouwe}, \citenamefont {Floriani}, \citenamefont {Katona},
  \citenamefont {Marguet} \emph {et~al.}}]{perez2025}%
  \BibitemOpen
  \bibfield  {author} {\bibinfo {author} {\bibfnamefont {E.}~\bibnamefont
  {Perez-Martin}}, \bibinfo {author} {\bibfnamefont {T.}~\bibnamefont
  {Beranger}}, \bibinfo {author} {\bibfnamefont {L.}~\bibnamefont {Bonnet}},
  \bibinfo {author} {\bibfnamefont {F.}~\bibnamefont {Teppe}}, \bibinfo
  {author} {\bibfnamefont {A.}~\bibnamefont {Lisauskas}}, \bibinfo {author}
  {\bibfnamefont {K.}~\bibnamefont {Ikamas}}, \bibinfo {author} {\bibfnamefont
  {E.}~\bibnamefont {Vrouwe}}, \bibinfo {author} {\bibfnamefont
  {E.}~\bibnamefont {Floriani}}, \bibinfo {author} {\bibfnamefont
  {G.}~\bibnamefont {Katona}}, \bibinfo {author} {\bibfnamefont
  {D.}~\bibnamefont {Marguet}}, \emph {et~al.},\ }\bibfield  {title} {\bibinfo
  {title} {Unveiling long-range forces in light-harvesting proteins: Pivotal
  roles of temperature and light},\ }\href@noop {} {\bibfield  {journal}
  {\bibinfo  {journal} {Science Advances}\ }\textbf {\bibinfo {volume} {11}},\
  \bibinfo {pages} {eadv0346} (\bibinfo {year} {2025})}\BibitemShut {NoStop}%
\bibitem [{\citenamefont {Fr{\"o}hlich}(1968)}]{frohlich1968long}%
  \BibitemOpen
  \bibfield  {author} {\bibinfo {author} {\bibfnamefont {H.}~\bibnamefont
  {Fr{\"o}hlich}},\ }\bibfield  {title} {\bibinfo {title} {Long-range coherence
  and energy storage in biological systems},\ }\href@noop {} {\bibfield
  {journal} {\bibinfo  {journal} {International Journal of Quantum Chemistry}\
  }\textbf {\bibinfo {volume} {2}},\ \bibinfo {pages} {641} (\bibinfo {year}
  {1968})}\BibitemShut {NoStop}%
\bibitem [{\citenamefont {Preto}\ \emph {et~al.}(2015)\citenamefont {Preto},
  \citenamefont {Pettini},\ and\ \citenamefont
  {Tuszynski}}]{preto2015possible}%
  \BibitemOpen
  \bibfield  {author} {\bibinfo {author} {\bibfnamefont {J.}~\bibnamefont
  {Preto}}, \bibinfo {author} {\bibfnamefont {M.}~\bibnamefont {Pettini}},\
  and\ \bibinfo {author} {\bibfnamefont {J.~A.}\ \bibnamefont {Tuszynski}},\
  }\bibfield  {title} {\bibinfo {title} {Possible role of electrodynamic
  interactions in long-distance biomolecular recognition},\ }\href@noop {}
  {\bibfield  {journal} {\bibinfo  {journal} {Physical Review E}\ }\textbf
  {\bibinfo {volume} {91}},\ \bibinfo {pages} {052710} (\bibinfo {year}
  {2015})}\BibitemShut {NoStop}%
\bibitem [{\citenamefont {Hameroff}(1998)}]{stuart1998quantum}%
  \BibitemOpen
  \bibfield  {author} {\bibinfo {author} {\bibfnamefont {S.}~\bibnamefont
  {Hameroff}},\ }\bibfield  {title} {\bibinfo {title} {Quantum computation in
  brain microtubules? the {Penrose-Hameroff} {'Orch OR'} model of
  consciousness},\ }\href@noop {} {\bibfield  {journal} {\bibinfo  {journal}
  {Philosophical Transactions of the Royal Society of London. Series A:
  Mathematical, Physical and Engineering Sciences}\ }\textbf {\bibinfo {volume}
  {356}},\ \bibinfo {pages} {1869} (\bibinfo {year} {1998})}\BibitemShut
  {NoStop}%
\bibitem [{\citenamefont {Reimers}\ \emph {et~al.}(2009)\citenamefont
  {Reimers}, \citenamefont {McKemmish}, \citenamefont {McKenzie}, \citenamefont
  {Mark},\ and\ \citenamefont {Hush}}]{reimers2009weak}%
  \BibitemOpen
  \bibfield  {author} {\bibinfo {author} {\bibfnamefont {J.~R.}\ \bibnamefont
  {Reimers}}, \bibinfo {author} {\bibfnamefont {L.~K.}\ \bibnamefont
  {McKemmish}}, \bibinfo {author} {\bibfnamefont {R.~H.}\ \bibnamefont
  {McKenzie}}, \bibinfo {author} {\bibfnamefont {A.~E.}\ \bibnamefont {Mark}},\
  and\ \bibinfo {author} {\bibfnamefont {N.~S.}\ \bibnamefont {Hush}},\
  }\bibfield  {title} {\bibinfo {title} {Weak, strong, and coherent regimes of
  {Fr\"ohlich} condensation and their applications to terahertz medicine and
  quantum consciousness},\ }\href@noop {} {\bibfield  {journal} {\bibinfo
  {journal} {PNAS}\ }\textbf {\bibinfo {volume} {106}},\ \bibinfo {pages}
  {4219} (\bibinfo {year} {2009})}\BibitemShut {NoStop}%
\bibitem [{\citenamefont {Azizi}\ \emph {et~al.}(2023)\citenamefont {Azizi},
  \citenamefont {Gori}, \citenamefont {Morzan}, \citenamefont {Hassanali},\
  and\ \citenamefont {Kurian}}]{azizi2023examining}%
  \BibitemOpen
  \bibfield  {author} {\bibinfo {author} {\bibfnamefont {K.}~\bibnamefont
  {Azizi}}, \bibinfo {author} {\bibfnamefont {M.}~\bibnamefont {Gori}},
  \bibinfo {author} {\bibfnamefont {U.}~\bibnamefont {Morzan}}, \bibinfo
  {author} {\bibfnamefont {A.}~\bibnamefont {Hassanali}},\ and\ \bibinfo
  {author} {\bibfnamefont {P.}~\bibnamefont {Kurian}},\ }\bibfield  {title}
  {\bibinfo {title} {Examining the origins of observed terahertz modes from an
  optically pumped atomistic model protein in aqueous solution},\ }\href@noop
  {} {\bibfield  {journal} {\bibinfo  {journal} {PNAS nexus}\ }\textbf
  {\bibinfo {volume} {2}},\ \bibinfo {pages} {pgad257} (\bibinfo {year}
  {2023})}\BibitemShut {NoStop}%
\bibitem [{\citenamefont {Tenenbaum}(2024)}]{tenenbaum2024energy}%
  \BibitemOpen
  \bibfield  {author} {\bibinfo {author} {\bibfnamefont {A.}~\bibnamefont
  {Tenenbaum}},\ }\bibfield  {title} {\bibinfo {title} {Energy condensation and
  dipole alignment in protein dynamics},\ }\href@noop {} {\bibfield  {journal}
  {\bibinfo  {journal} {Physical Review E}\ }\textbf {\bibinfo {volume}
  {109}},\ \bibinfo {pages} {044401} (\bibinfo {year} {2024})}\BibitemShut
  {NoStop}%
\bibitem [{\citenamefont {Kolossv{\'a}ry}(2024)}]{kolossvary2024}%
  \BibitemOpen
  \bibfield  {author} {\bibinfo {author} {\bibfnamefont {I.}~\bibnamefont
  {Kolossv{\'a}ry}},\ }\bibfield  {title} {\bibinfo {title} {A fresh look at
  the normal mode analysis of proteins: Introducing allosteric co-vibrational
  modes},\ }\href@noop {} {\bibfield  {journal} {\bibinfo  {journal} {JACS Au}\
  }\textbf {\bibinfo {volume} {4}},\ \bibinfo {pages} {1303} (\bibinfo {year}
  {2024})}\BibitemShut {NoStop}%
\bibitem [{\citenamefont {Moritsugu}\ \emph {et~al.}(2000)\citenamefont
  {Moritsugu}, \citenamefont {Miyashita},\ and\ \citenamefont
  {Kidera}}]{moritsugu2000}%
  \BibitemOpen
  \bibfield  {author} {\bibinfo {author} {\bibfnamefont {K.}~\bibnamefont
  {Moritsugu}}, \bibinfo {author} {\bibfnamefont {O.}~\bibnamefont
  {Miyashita}},\ and\ \bibinfo {author} {\bibfnamefont {A.}~\bibnamefont
  {Kidera}},\ }\bibfield  {title} {\bibinfo {title} {Vibrational energy
  transfer in a protein molecule},\ }\href@noop {} {\bibfield  {journal}
  {\bibinfo  {journal} {Physical review letters}\ }\textbf {\bibinfo {volume}
  {85}},\ \bibinfo {pages} {3970} (\bibinfo {year} {2000})}\BibitemShut
  {NoStop}%
\bibitem [{\citenamefont {Go}\ \emph {et~al.}(1983)\citenamefont {Go},
  \citenamefont {Noguti},\ and\ \citenamefont {Nishikawa}}]{go1983dynamics}%
  \BibitemOpen
  \bibfield  {author} {\bibinfo {author} {\bibfnamefont {N.}~\bibnamefont
  {Go}}, \bibinfo {author} {\bibfnamefont {T.}~\bibnamefont {Noguti}},\ and\
  \bibinfo {author} {\bibfnamefont {T.}~\bibnamefont {Nishikawa}},\ }\bibfield
  {title} {\bibinfo {title} {Dynamics of a small globular protein in terms of
  low-frequency vibrational modes},\ }\href@noop {} {\bibfield  {journal}
  {\bibinfo  {journal} {PNAS}\ }\textbf {\bibinfo {volume} {80}},\ \bibinfo
  {pages} {3696} (\bibinfo {year} {1983})}\BibitemShut {NoStop}%
\bibitem [{\citenamefont {Preto}(2016)}]{preto2016classical}%
  \BibitemOpen
  \bibfield  {author} {\bibinfo {author} {\bibfnamefont {J.}~\bibnamefont
  {Preto}},\ }\bibfield  {title} {\bibinfo {title} {Classical investigation of
  long-range coherence in biological systems},\ }\href@noop {} {\bibfield
  {journal} {\bibinfo  {journal} {Chaos: An Interdisciplinary Journal of
  Nonlinear Science}\ }\textbf {\bibinfo {volume} {26}} (\bibinfo {year}
  {2016})}\BibitemShut {NoStop}%
\bibitem [{sup()}]{suppmat}%
  \BibitemOpen
  \href@noop {} {\bibinfo {title} {See {Supplemental Material} at [{URL}] for
  additional figures, tables, and comments on the original {FREs postulated by
  Fr\"ohlich}.}}\BibitemShut {Stop}%
\bibitem [{\citenamefont {Wu}\ and\ \citenamefont {Austin}(1977)}]{wu1977bose}%
  \BibitemOpen
  \bibfield  {author} {\bibinfo {author} {\bibfnamefont {T.}~\bibnamefont
  {Wu}}\ and\ \bibinfo {author} {\bibfnamefont {S.}~\bibnamefont {Austin}},\
  }\bibfield  {title} {\bibinfo {title} {Bose condensation in biosystems},\
  }\href@noop {} {\bibfield  {journal} {\bibinfo  {journal} {Physics Letters
  A}\ }\textbf {\bibinfo {volume} {64}},\ \bibinfo {pages} {151} (\bibinfo
  {year} {1977})}\BibitemShut {NoStop}%
\bibitem [{\citenamefont {Wu}\ and\ \citenamefont
  {Austin}(1981)}]{wu1981frohlich}%
  \BibitemOpen
  \bibfield  {author} {\bibinfo {author} {\bibfnamefont {T.}~\bibnamefont
  {Wu}}\ and\ \bibinfo {author} {\bibfnamefont {S.~J.}\ \bibnamefont
  {Austin}},\ }\bibfield  {title} {\bibinfo {title} {Fr{\"o}hlich's model of
  bose condensation in biological systems},\ }\href@noop {} {\bibfield
  {journal} {\bibinfo  {journal} {Journal of Biological Physics}\ }\textbf
  {\bibinfo {volume} {9}},\ \bibinfo {pages} {97} (\bibinfo {year}
  {1981})}\BibitemShut {NoStop}%
\bibitem [{\citenamefont {Mills}(1983)}]{mills1983frohlich}%
  \BibitemOpen
  \bibfield  {author} {\bibinfo {author} {\bibfnamefont {R.~E.}\ \bibnamefont
  {Mills}},\ }\bibfield  {title} {\bibinfo {title} {Fr{\"o}hlich's model of
  nonthermal excitations in biological systems},\ }\href@noop {} {\bibfield
  {journal} {\bibinfo  {journal} {Physical Review A}\ }\textbf {\bibinfo
  {volume} {28}},\ \bibinfo {pages} {379} (\bibinfo {year} {1983})}\BibitemShut
  {NoStop}%
\bibitem [{\citenamefont {Zwanzig}\ \emph {et~al.}(1972)\citenamefont
  {Zwanzig}, \citenamefont {Nordholm},\ and\ \citenamefont
  {Mitchell}}]{zwanzig1972memory}%
  \BibitemOpen
  \bibfield  {author} {\bibinfo {author} {\bibfnamefont {R.}~\bibnamefont
  {Zwanzig}}, \bibinfo {author} {\bibfnamefont {K.}~\bibnamefont {Nordholm}},\
  and\ \bibinfo {author} {\bibfnamefont {W.}~\bibnamefont {Mitchell}},\
  }\bibfield  {title} {\bibinfo {title} {Memory effects in irreversible
  thermodynamics: Corrected derivation of transport equations},\ }\href@noop {}
  {\bibfield  {journal} {\bibinfo  {journal} {Physical Review A}\ }\textbf
  {\bibinfo {volume} {5}},\ \bibinfo {pages} {2680} (\bibinfo {year}
  {1972})}\BibitemShut {NoStop}%
\bibitem [{\citenamefont {Nordholm}\ and\ \citenamefont
  {Zwanzig}(1975)}]{nordholm1975systematic}%
  \BibitemOpen
  \bibfield  {author} {\bibinfo {author} {\bibfnamefont {S.}~\bibnamefont
  {Nordholm}}\ and\ \bibinfo {author} {\bibfnamefont {R.}~\bibnamefont
  {Zwanzig}},\ }\bibfield  {title} {\bibinfo {title} {A systematic derivation
  of exact generalized brownian motion theory},\ }\href@noop {} {\bibfield
  {journal} {\bibinfo  {journal} {Journal of Statistical Physics}\ }\textbf
  {\bibinfo {volume} {13}},\ \bibinfo {pages} {347} (\bibinfo {year}
  {1975})}\BibitemShut {NoStop}%
\bibitem [{\citenamefont {Burgie}\ \emph {et~al.}(2014)\citenamefont {Burgie},
  \citenamefont {Wang}, \citenamefont {Bussell}, \citenamefont {Walker},
  \citenamefont {Li},\ and\ \citenamefont {Vierstra}}]{Burgie2014}%
  \BibitemOpen
  \bibfield  {author} {\bibinfo {author} {\bibfnamefont {E.~S.}\ \bibnamefont
  {Burgie}}, \bibinfo {author} {\bibfnamefont {T.}~\bibnamefont {Wang}},
  \bibinfo {author} {\bibfnamefont {A.~N.}\ \bibnamefont {Bussell}}, \bibinfo
  {author} {\bibfnamefont {J.~M.}\ \bibnamefont {Walker}}, \bibinfo {author}
  {\bibfnamefont {H.}~\bibnamefont {Li}},\ and\ \bibinfo {author}
  {\bibfnamefont {R.~D.}\ \bibnamefont {Vierstra}},\ }\bibfield  {title}
  {\bibinfo {title} {Crystallographic and electron microscopic analyses of a
  bacterial phytochrome reveal local and global rearrangements during
  photoconversion},\ }\href@noop {} {\bibfield  {journal} {\bibinfo  {journal}
  {Journal of Biological Chemistry}\ }\textbf {\bibinfo {volume} {289}},\
  \bibinfo {pages} {24573} (\bibinfo {year} {2014})}\BibitemShut {NoStop}%
\bibitem [{\citenamefont {Poddar}\ \emph {et~al.}(2022)\citenamefont {Poddar},
  \citenamefont {Heyes}, \citenamefont {Schir{\`o}}, \citenamefont {Weik},
  \citenamefont {Leys},\ and\ \citenamefont {Scrutton}}]{Poddar2022}%
  \BibitemOpen
  \bibfield  {author} {\bibinfo {author} {\bibfnamefont {H.}~\bibnamefont
  {Poddar}}, \bibinfo {author} {\bibfnamefont {D.~J.}\ \bibnamefont {Heyes}},
  \bibinfo {author} {\bibfnamefont {G.}~\bibnamefont {Schir{\`o}}}, \bibinfo
  {author} {\bibfnamefont {M.}~\bibnamefont {Weik}}, \bibinfo {author}
  {\bibfnamefont {D.}~\bibnamefont {Leys}},\ and\ \bibinfo {author}
  {\bibfnamefont {N.~S.}\ \bibnamefont {Scrutton}},\ }\bibfield  {title}
  {\bibinfo {title} {A guide to time-resolved structural analysis of
  light-activated proteins},\ }\href@noop {} {\bibfield  {journal} {\bibinfo
  {journal} {The FEBS Journal}\ }\textbf {\bibinfo {volume} {289}},\ \bibinfo
  {pages} {576} (\bibinfo {year} {2022})}\BibitemShut {NoStop}%
\bibitem [{\citenamefont {Legris}\ \emph {et~al.}(2016)\citenamefont {Legris},
  \citenamefont {Klose}, \citenamefont {Burgie}, \citenamefont {Rojas},
  \citenamefont {Neme}, \citenamefont {Hiltbrunner}, \citenamefont {Wigge},
  \citenamefont {Sch{\"a}fer}, \citenamefont {Vierstra},\ and\ \citenamefont
  {Casal}}]{Legris2016}%
  \BibitemOpen
  \bibfield  {author} {\bibinfo {author} {\bibfnamefont {M.}~\bibnamefont
  {Legris}}, \bibinfo {author} {\bibfnamefont {C.}~\bibnamefont {Klose}},
  \bibinfo {author} {\bibfnamefont {E.~S.}\ \bibnamefont {Burgie}}, \bibinfo
  {author} {\bibfnamefont {C.~C.~R.}\ \bibnamefont {Rojas}}, \bibinfo {author}
  {\bibfnamefont {M.}~\bibnamefont {Neme}}, \bibinfo {author} {\bibfnamefont
  {A.}~\bibnamefont {Hiltbrunner}}, \bibinfo {author} {\bibfnamefont {P.~A.}\
  \bibnamefont {Wigge}}, \bibinfo {author} {\bibfnamefont {E.}~\bibnamefont
  {Sch{\"a}fer}}, \bibinfo {author} {\bibfnamefont {R.~D.}\ \bibnamefont
  {Vierstra}},\ and\ \bibinfo {author} {\bibfnamefont {J.~J.}\ \bibnamefont
  {Casal}},\ }\bibfield  {title} {\bibinfo {title} {Phytochrome {B} integrates
  light and temperature signals in arabidopsis},\ }\href@noop {} {\bibfield
  {journal} {\bibinfo  {journal} {Science}\ }\textbf {\bibinfo {volume}
  {354}},\ \bibinfo {pages} {897} (\bibinfo {year} {2016})}\BibitemShut
  {NoStop}%
\bibitem [{\citenamefont {Lifshits}(1972)}]{Lifshits1972problem}%
  \BibitemOpen
  \bibfield  {author} {\bibinfo {author} {\bibfnamefont {M.~A.}\ \bibnamefont
  {Lifshits}},\ }\bibfield  {title} {\bibinfo {title} {Problem of the
  participation of coherent phonons in biological processes},\ }\href@noop {}
  {\bibfield  {journal} {\bibinfo  {journal} {Biophysics (USSR)}\ }\textbf
  {\bibinfo {volume} {17}},\ \bibinfo {pages} {726} (\bibinfo {year}
  {1972})}\BibitemShut {NoStop}%
\bibitem [{\citenamefont {Moritsugu}\ \emph {et~al.}(2003)\citenamefont
  {Moritsugu}, \citenamefont {Miyashita},\ and\ \citenamefont
  {Kidera}}]{moritsugu2003temperature}%
  \BibitemOpen
  \bibfield  {author} {\bibinfo {author} {\bibfnamefont {K.}~\bibnamefont
  {Moritsugu}}, \bibinfo {author} {\bibfnamefont {O.}~\bibnamefont
  {Miyashita}},\ and\ \bibinfo {author} {\bibfnamefont {A.}~\bibnamefont
  {Kidera}},\ }\bibfield  {title} {\bibinfo {title} {Temperature dependence of
  vibrational energy transfer in a protein molecule},\ }\href@noop {}
  {\bibfield  {journal} {\bibinfo  {journal} {The Journal of Physical Chemistry
  B}\ }\textbf {\bibinfo {volume} {107}},\ \bibinfo {pages} {3309} (\bibinfo
  {year} {2003})}\BibitemShut {NoStop}%
\bibitem [{\citenamefont {Pandey}\ and\ \citenamefont
  {Cifra}(2024)}]{pandey2024tubulin}%
  \BibitemOpen
  \bibfield  {author} {\bibinfo {author} {\bibfnamefont {S.~K.}\ \bibnamefont
  {Pandey}}\ and\ \bibinfo {author} {\bibfnamefont {M.}~\bibnamefont {Cifra}},\
  }\bibfield  {title} {\bibinfo {title} {Tubulin vibration modes are in the
  subterahertz range, and their electromagnetic absorption is affected by
  water},\ }\href@noop {} {\bibfield  {journal} {\bibinfo  {journal} {The
  Journal of Physical Chemistry Letters}\ }\textbf {\bibinfo {volume} {15}},\
  \bibinfo {pages} {8334} (\bibinfo {year} {2024})}\BibitemShut {NoStop}%
\bibitem [{\citenamefont {Bauer}\ \emph {et~al.}(2019)\citenamefont {Bauer},
  \citenamefont {Pavlovi{\'c}},\ and\ \citenamefont
  {Bauerov{\'a}-Hlinkov{\'a}}}]{bauer2019normal}%
  \BibitemOpen
  \bibfield  {author} {\bibinfo {author} {\bibfnamefont {J.~A.}\ \bibnamefont
  {Bauer}}, \bibinfo {author} {\bibfnamefont {J.}~\bibnamefont
  {Pavlovi{\'c}}},\ and\ \bibinfo {author} {\bibfnamefont {V.}~\bibnamefont
  {Bauerov{\'a}-Hlinkov{\'a}}},\ }\bibfield  {title} {\bibinfo {title} {Normal
  mode analysis as a routine part of a structural investigation},\ }\href@noop
  {} {\bibfield  {journal} {\bibinfo  {journal} {Molecules}\ }\textbf {\bibinfo
  {volume} {24}},\ \bibinfo {pages} {3293} (\bibinfo {year}
  {2019})}\BibitemShut {NoStop}%
\end{thebibliography}
\end{document}